\documentstyle[onecolumn]{mn}

\newif\ifAMStwofonts
\ifoldfss
  \ifCUPmtlplainloaded \else
    \NewTextAlphabet{textbfit} {cmbxti10} {}
    \NewTextAlphabet{textbfss} {cmssbx10} {}
    \NewMathAlphabet{mathbfit} {cmbxti10} {} % for math mode
    \NewMathAlphabet{mathbfss} {cmssbx10} {} %  "   "    "
  \fi
  \ifAMStwofonts
    \ifCUPmtlplainloaded \else
      \NewSymbolFont{upmath} {eurm10}
      \NewSymbolFont{AMSa} {msam10}
      \NewMathSymbol{\upi}     {0}{upmath}{19}
      \NewMathSymbol{\umu}     {0}{upmath}{16}
      \NewMathSymbol{\upartial}{0}{upmath}{40}
      \NewMathSymbol{\leqslant}{3}{AMSa}{36}
      \NewMathSymbol{\geqslant}{3}{AMSa}{3E}

      \let\geq=\geqslant 
    \fi
  \fi
\fi % End of OFSS

\ifnfssone
  \newmathalphabet{\mathit}
  \addtoversion{normal}{\mathit}{cmr}{m}{it}
  \addtoversion{bold}{\mathit}{cmr}{bx}{it}
  \newmathalphabet{\mathbfit} % math mode version of \textbfit{..}
  \addtoversion{normal}{\mathbfit}{cmr}{bx}{it}
  \addtoversion{bold}{\mathbfit}{cmr}{bx}{it}
  \newmathalphabet{\mathbfss} % math mode version of \textbfss{..}
  \addtoversion{normal}{\mathbfss}{cmss}{bx}{n}
  \addtoversion{bold}{\mathbfss}{cmss}{bx}{n}
  \ifAMStwofonts
    \ifCUPmtlplainloaded \else
      \UseAMStwoboldmath
      \makeatletter
      \new@mathgroup\upmath@group
      \define@mathgroup\mv@normal\upmath@group{eur}{m}{n}
      \define@mathgroup\mv@bold\upmath@group{eur}{b}{n}
      \edef\UPM{\hexnumber\upmath@group}
      \new@mathgroup\amsa@group
      \define@mathgroup\mv@normal\amsa@group{msa}{m}{n}
      \define@mathgroup\mv@bold\amsa@group{msa}{m}{n}
      \edef\AMSa{\hexnumber\amsa@group}
      \makeatother
      \mathchardef\upi="0\UPM19
      \mathchardef\umu="0\UPM16
      \mathchardef\upartial="0\UPM40
      \mathchardef\leqslant="3\AMSa36
      \mathchardef\geqslant="3\AMSa3E

      \let\geq=\geqslant 
    \fi
  \fi
\fi % End of NFSS release 1

\ifnfsstwo
  \DeclareMathAlphabet{\mathbfit}{OT1}{cmr}{bx}{it}
  \SetMathAlphabet\mathbfit{bold}{OT1}{cmr}{bx}{it}
  \DeclareMathAlphabet{\mathbfss}{OT1}{cmss}{bx}{n}
  \SetMathAlphabet\mathbfss{bold}{OT1}{cmss}{bx}{n}
  \ifAMStwofonts
    \ifCUPmtlplainloaded \else
      \DeclareSymbolFont{UPM}{U}{eur}{m}{n}
      \SetSymbolFont{UPM}{bold}{U}{eur}{b}{n}
      \DeclareSymbolFont{AMSa}{U}{msa}{m}{n}
      \DeclareMathSymbol{\upi}{0}{UPM}{"19}
      \DeclareMathSymbol{\umu}{0}{UPM}{"16}
      \DeclareMathSymbol{\upartial}{0}{UPM}{"40}
      \DeclareMathSymbol{\leqslant}{3}{AMSa}{"36}
      \DeclareMathSymbol{\geqslant}{3}{AMSa}{"3E}

      \let\geq=\geqslant 
    \fi
  \fi
\fi % End of NFSS release 2

\ifCUPmtlplainloaded \else
  \ifAMStwofonts \else % If no AMS fonts
    \def\upi{\pi}
    \def\umu{\mu}
    \def\upartial{\partial}
  \fi
\fi

\newcommand{\beq}{\begin{equation}}
\newcommand{\beqa}{\begin{eqnarray}}
\newcommand{\eeq}{\end{equation}}
\newcommand{\eeqa}{\end{eqnarray}}
\newcommand{\be}{\begin{equation}}
\newcommand{\ba}{\begin{eqnarray}}
\newcommand{\ee}{\end{equation}}
\newcommand{\ea}{\end{eqnarray}}

\newcommand{\k}{\kappa}

\journal{YITP-98-6(January 1998),astro-ph/9810447}

\title[Determining the Equation of State of the Universe]{
Determining the Equation of State of the Expanding Universe:
Inverse Problem in Cosmology    }

\author[Nakamura and Chiba]{Takashi Nakamura$^1$ and  
Takeshi Chiba$^2$\\
$^1$Yukawa Institute for Theoretical Physics, Kyoto University, 
Kyoto 606-8502, Japan\\
$^2$Department of Physics, University of
Tokyo, Tokyo 113-0033, Japan
}

\date{Accepted .
      Received ;
      in original form 1998 January 19}

\pubyear{1998}

\begin{document}

\maketitle

%\label{firstpage}

\begin{abstract}
Even if the luminosity distance as a function of redshift is obtained 
accurately using, for example, Type Ia supernovae, the equation of
state of the Universe cannot be determined uniquely but 
depends on one free parameter $\Omega_{k0} ={k}/(a_0^2H_0^2)$, 
where $a_0$ and $H_0$ are the present scale factor and the Hubble
parameter, respectively. This degeneracy might be resolved if, for
example, the time variations  of the redshift of quasars are measured
as proposed recently by Loeb. Therefore the equation of state of the
Universe (or the metric of the universe) might be determined 
without any theoretical assumption on the 
matter content of the Universe in future.

\end{abstract}
   

\section{introduction}

 To determine the structure and  dynamics of an astrophysical system,
the equation of state is usually necessary. For example, consider the
structure of a spherical neutron star.  If the pressure $p$ is known 
as a function of the density $\rho$, we can determine the gravitational
mass $M$ and the radius $R$ of the star as a function of the central density
$\rho_c$ by solving the Oppenheimer-Volkoff equation \cite{ov}. This
means that we can determine the mass-radius relation $M(R)$ 
theoretically in principle.  However, the
equation of state relevant to the neutron star is not established yet, 
although it may be determined by Quantum Chromo Dynamics in future.
Therefore the mass-radius relation of the neutron star is not 
known well theoretically at present.

  Observationally, however, the mass-radius relation of neutron stars 
may be determined in the near future. If gravitational waves from the
coalescing binary neutron stars are detected by the LIGO/VIRGO/GEO/TAMA
network which will be in operation around 2000, the  mass of each neutron
star  as well as its radius  may be determined by analyzing the
waveform in the last three minutes of the binary inspiral \cite{kip}.
In general the mass of each observed neutron star can be different 
so that there is a chance to determine $M(R)$ observationally. 
In this case,  as shown by Lindblom \shortcite{lindblom}, 
the  equation of state of the high
density matter can be determined from $M(R)$. This is, in a sense, the
inverse problem.

Now let us consider an isotropic and homogeneous universe which
describes the Universe quite well in a global sense. 
The amount of the radiation $\rho_r(z)$ in the Universe is well known
as a function of the redshift $z$ from the present temperature of the
cosmic background radiation. 
The amount of baryons $\rho_b(z)$ has a constraint from the big bang 
nucleosynthesis. 
We know that dark matter should exist but we do not know well its
density $\rho_d(z)$ or pressure $p_d(z)$. A cosmological constant 
may exist but introducing a non-zero cosmological constant
needs a fine-tuning of the vacuum energy, and at present we do not
have any convincing explanation for  
why such an extremely small value of the
cosmological constant (in Planck units) is required. 
Since we do not know the equation of state of
the  matter in the standard model well, we cannot determine the scale factor as a
function of  time theoretically.  Moreover recently several authors
have considered  the more  
general equation of state for a dark component called x-matter or
quintessence and have explored its
cosmological implications \cite{tw,xmatter,xmatter2,qmatter,lumi}.
The  situation is worse  than the neutron star case; 
the equation of state in the expanding universe is almost unknown
theoretically.

Observationally we have several quantities such as the luminosity
distance, the angular diameter distance and number counts  as a
function of the redshift. Among these the luminosity distance  $d_L(z)$ 
 may be determined quite accurately by using Type Ia 
supernovae\cite{sn1,sn2,sn3,sn4}. In the future the afterglows of a certain 
class of gamma ray bursts  might serve as  
a standard candle\cite{piran}. Therefore in this paper we assume that
a quite accurate luminosity distance  may be obtained and
examine whether the equation of state of the expanding universe can be
determined uniquely. Namely we discuss the inverse problem in
cosmology. By ``equation of state of the universe'', we mean the
relation between the total energy density of cosmic matter and the
total pressure. 
Nearly three decades ago, Weinberg studied the possibility
of determining the metric from the observed luminosity distance, 
with negative conclusion \cite{weinberg}. We shall argue that 
the improvement of observational techniques now enables us to 
determine the metric of the universe (or the equation of the
state of the universe) directly from the
observational data. 

In section 2 we show that the equation of state can be
determined if the scale factor is given as a function of time. 
We also show that 
the scalar field potential can be determined similarly. 
In section 3 we discuss how to determine the metric of the universe
{} from the luminosity distance as a function of the redshift $z$.

\section{ matter field in terms of scale factor}

Consider any given  scale factor $a(t)$ which is  a monotonically
increasing function of time. We denote the inverse
function of $a(t)$ as $t(a)$. Then  every function of $t$ can be
considered as 
a function of $a$. For example the Hubble parameter $H$ can be
written as 
\beq
\frac{\dot a}{a}=\left(a\frac{dt}{da}\right)^{-1}=H(a)
\eeq
The metric of our isotropic and homogeneous universe is given by
\be
ds^2=-dt^2+a(t)^2(d\chi^2+f(\chi)^2d\Omega^2)
\ee
where $f(\chi)=\chi, \sinh (\chi)$ and $\sin (\chi)$ for flat, open
and closed universes, respectively. From the general
classification of energy momentum tensors \cite{he},
the form of the energy momentum tensor compatible with the metric (2)
is of Type I with pressure $p_1=p_2=p_3=p(a)$.  We express the total
energy density  as 
$\rho(a)$. Then $p$ can be expressed by $\rho$ as $p(\rho)$ which we
call the equation of state of the Universe. The Einstein equations are 
\beqa
H(a)^2&=& {\k^2 \over3}\rho-{k\over a^2},\label{fluid:rho}\\
\dot H(a) +H(a)^2&=& -{\k^2 \over 6}\left(\rho+3p\right),\label{fluid:p}\\
\dot \rho &=& -3H\left(\rho+p\right),
\eeqa
where $\k^2=8\pi G$ and $k=0, -1$ and  $1$ for flat, open and closed 
universes, respectively.
First, from Eq.(\ref{fluid:rho}) $\rho$ as a function
of $a$ can be written   as
\beq
\k^2\rho(a)={3}\left(H(a)^2+{k\over a^2}\right).
\eeq
Then, from Eq.(\ref{fluid:p}) $p$ as a function
of $a$ can also be written  as
\beq
\k^2p(a)=-\left(2\dot H(a) +3H(a)^2 +{k\over a^2}\right).
\eeq

Since $\rho(a)$ and $p(a)$ are given, we can determine the equation of
state $p(\rho)$. Note that  we can
consider even the case where the weak energy condition \cite{he} 
is violated: $\rho+p
< 0$ or $\rho <0$.

 Next consider the minimally coupled scalar field $\phi$ with the potential
$V(\phi)$, which is an 
example of the general energy momentum tensor described above.
Then
\beqa
&&\rho=\frac{1}{2}\dot \phi^2+V(\phi),\label{rho}\\
&&p=\frac{1}{2}\dot \phi^2-V(\phi),\label{p}\\
&&\ddot \phi+ 3H\dot \phi = -V'.
\eeqa
In this case the weak energy condition is satisfied.
We assume $ \dot \phi \geq 0$  for simplicity though it  is not necessary.  
{}From Eqs.(\ref{rho}) and (\ref{p})  we immediately have,
\beqa
&&\k^2V=\left(3H^2+\dot H +{2k\over
a^2}\right),\label{poten}\\
&&\k^2\dot \phi^2=\k^2\left({d\phi\over da}Ha\right)^2=
 -2\left(\dot H - {k\over
a^2}\right).\label{kinetic}
\eeqa
{}From Eq.(\ref{poten}) $V$ can be
written as a function of $a$. Also, from Eq.(\ref{kinetic}) $\phi$ can 
be written as a function of $a$. Therefore, $V$ can be written as a
function of $\phi$. 

\section{ determining the metric from 
observations}

\subsection{Matter field in terms of the luminosity distance}
{}From observations we might know, 
for example, the luminosity distance as a 
function of redshift $d_L(z)$ accurately  by using  Type Ia
supernovae\cite{sn1,sn2,sn3,sn4}. 
In this section  we therefore will regard  the Hubble parameter $H(z)$
as a function of the redshift instead of as a function of the scale
factor in the previous section.
 $\rho(z)$ and $p(z)$ are then given by
\ba
&&\k^2\rho(z) ={3}\left(H(z)^2+(1+z)^2H_0^2\Omega_{k0}\right), \\
&&\k^2p(z)    = -3H(z)^2+2(1+z)H(z)\frac{dH}{dz}-
(1+z)^2H_0^2\Omega_{k0},\\
&&\Omega_{k0} \equiv \frac{k}{a_0^2H_0^2},
\ea
where $a_0$ and $H_0$ are the present scale factor and the Hubble
parameter, respectively. It  is apparent that in general 
the equation of state of the Universe depends on $\Omega_{k0}$.
To determine $H(z)$ we will use  the luminosity distance. Since all  
cosmological
observations are made on the past light cone, the argument is
similar for other distance indicators.
The luminosity distance $d_L(z)$ is defined by 
\beqa
&&d_L(z)=a_0(1+z)f(\chi)\equiv (1+z)r(z),\\
&&\chi={1\over a_0}\int^z_0{du\over H(u)}.
\eeqa
Then $H(z)$ can be written in terms of $r(z)$ as 
\beq
H(z)=\left(dr/dz \right)^{-1}
\sqrt{1- r(z)^2H_0^2\Omega_{k0}}.
\label{hubble}
\eeq
Note here that the above formula is valid irrespective of the sign of
$k$. Since $r(0)=0$, $H_0$ can be determined
irrespective of $\Omega_{k0}$.  One may think that  $\Omega_{k0}$ can 
be determined only from $r(z)$. However, this is not the case. To show 
this, we rewrite Eq.(\ref{hubble}) as
\be
{dr\over dz} =H(z)^{-1}
\sqrt{1- r(z)^2H_0^2\Omega_{k0}}.
\label{dlprime}
\ee
{}From the second derivative of Eq.(\ref{dlprime}),
we have
\be
{d^3 r\over dz^3}(0)={d^2 \over dz^2}
\left(\frac{1}{H}\right)_0 -\frac{\Omega_{k0}}{H_0}.
\label{dlprimet}
\ee
Eq.(\ref{dlprimet}) shows that we cannot determine  $\Omega_{k0}$
without the knowledge of the second derivative of the Hubble
parameter. This is due to the fact that as far as the expansion
of the universe is concerned, the effect of the curvature is
equivalent to ``matter'' with the equation of state 
$p=-\rho/3$.

Now using $r(z)$ we express $\rho(z)$ and $p(z)$ explicitly as
\beqa
&&\k^2\rho(z) =3\left[ 
 {1\over \left(dr/dz\right)^2}
+\left((1+z)^2-\frac{r^2}{\left(dr/dz\right)^2}
\right)H_0^2 \Omega_{k0}\right],\label{rhoz}\\
&&\k^2p(z) = -
 {3\over \left(dr/dz\right)^2}+(1+z){d\over dz}
\left({1\over (dr/dz)^2}\right)
- \left[(1+z)^2-{3r^2\over \left(dr/dz\right)^2}+
(1+z){d\over dz}\left({r^2\over 
(dr/dz)^2}\right)\right]H_0^2\Omega_{k0} .\label{pz}
\eeqa
The above equations show that  the equation of state
of the Universe cannot be determined uniquely from the luminosity 
distance $d_L(z)=(1+z)r(z)$ but depends on one free parameter
$\Omega_{k0}$. This degeneracy was first pointed out by
Weinberg\shortcite{weinberg}. 
This means that even if the luminosity distance
is observed accurately as a function of $z$, in order 
to determine the equation of the state of the Universe (or the metric) 
some assumption about 
$\Omega_{k0}$ is needed. Intuitively, this can be understood as
follows. Since the luminosity distance or other cosmological distance
measure only  gives information  on the past light cone, it is
not enough to infer what happens   inside the light cone. 

\subsection{Breaking the degeneracy}

To determine $\Omega_{k0}$  we need other dynamical 
information.   We here show that $\Omega_{k0}$ can be determined if we 
use a new observational technique  proposed by
Loeb\shortcite{loeb}. He pointed out that the time variation
 of cosmic redshift  might be  detectable 
through two observations of $\sim 10^2$ quasars set a decade apart 
with the HIRES instrument of the Keck 10 meter telescope. The key
point of his proposal is to use an existing spectroscopic technique, 
which was  recently employed in planet searches. If a jupiter size
planet exists around a certain G/F type star, the G/F type
star is moving around the center of mass of the star-planet system
 with rotational velocity of the order of $\sim 30$m/s.
In principle, the wave length of an absorption line in the spectrum of
the star is doppler shifted. By observing only one absorption line it is 
difficult to determine a period and an orbital velocity to obtain 
the mass and the orbital radius of the planet because the expected shift of the
line is too small compared with the line width. However if one
observes many absorption lines, the signal-to-noise ratio(S/N) will
increase in proportion to $\sqrt{N}$ where
$N$ is the number of lines observed. The same technique can be
applied to cosmology in principle. Each line from a quasar has a large
line width so that it is difficult to determine the time variation of
the redshift. However by observing many quasars and  many absorption
lines it might be possible to determine one parameter $\Omega_{k0}$.
If such an observation is performed, the change of the 
redshift $\Delta z$  can be obtained as 
\beq
\Delta z =\left[(1+z)H_0-H(z)\right]\Delta t \equiv g(z)\Delta t
\eeq
where $\Delta t $  is the time interval of two observations. 
A spectroscopic velocity shift is then 
\beq
\Delta v={\Delta z\over 1+z}={g(z)\Delta t\over 1+z},
\eeq
which is of order $\sim 1 {\rm ms^{-1}}$ over $\Delta t \sim 10^2$
years for a single quasar. Loeb showed that it is feasible to detect 
the cosmic signal (with signal-to-noise ratio of $\sim$ 100) only 
over a decade with $\sim 10^2$ quasars. 

Since $g(z)$ can be determined observationally, $H(z)$ can be 
determined independently from Eq.(\ref{hubble}) as
\beq
H(z)=(1+z)H_0-g(z)
\eeq
In reality it may be hard to determine $g(z)$ for various $z$. 
However  if  $g(z)$ is determined at a  certain $z_s$ quite accurately, 
we can determine  $\Omega_{k0}$ as
\beq
\Omega_{k0}=r(z_s)^{-2}H_0^{-2}\left[1-\left(dr(z_s)/dz\right)^2
\left(\left(1+z_s\right)H_0-g\left(z_s\right)\right)^2\right]
\eeq
Even if $g(z)$ is not determined quite accurately for any $z$, if the
luminosity distance $d_L(z)=(1+z)r(z)$
 is obtained accurately, we may determine
$\Omega_{k0}$ by using a maximum likelihood test for many observations
of $g(z)$, since  only one parameter $\Omega_{k0}$ is to be determined. 

\subsection{Application to two component model}

Finally, we note that Eq.(\ref{rhoz}) and Eq.(\ref{pz}) can be 
applied to the two component model such that $\rho=\rho_M +\rho_X$,
where $\rho_M$ is dust matter and $\rho_X$ refers to the ``x-component'',
and that we can determine, once $\Omega_{M0}$ is known, the
equation of state (or the effective potential) of the x-component. 
For example, in the case of a fluid we have
\beqa
&&\k^2\rho_X(z) =3\left[ 
 {1\over \left(dr/dz\right)^2}
+\left((1+z)^2-\frac{r^2}{\left(dr/dz\right)^2}
\right)H_0^2 \Omega_{k0}-H_0^2\Omega_{M0}\right],\\
&&\k^2p_X(z) = -
 {3\over \left(dr/dz\right)^2}+(1+z){d\over dz}
\left({1\over (dr/dz)^2}\right)
- \left[(1+z)^2-{3r^2\over \left(dr/dz\right)^2}+
(1+z){d\over dz}\left({r^2\over 
(dr/dz)^2}\right)\right]H_0^2\Omega_{k0}.
\eeqa
Note that the cosmological constant corresponds to the $p_X=-\rho_X$ case.
Alternatively, in the case of a scalar field we have
\beqa
\k^2\dot\phi^2&=&\k^2(1+z)^2\left({1-r^2H_0^2\Omega_{k0}\over
      (dr/dz)^2}\right)\left({d\phi (z)\over dz}\right)^2 \nonumber\\
&=&(1+z){d\over dz}\left({1\over (dr/dz)^2}\right)+
\left[ 2(1+z)^2-(1+z){d\over dz}\left({r^2\over 
(dr/dz)^2}\right)\right]H_0^2\Omega_{k0}
-3H_0^2\Omega_{M0}\\
\k^2V(z)&=&{3\over \left(dr/dz\right)^2}
-{1\over 2}(1+z){d\over dz}
\left({1\over (dr/dz)^2}\right)+ \left[
2(1+z)^2-{3r^2\over \left(dr/dz\right)^2}+
{1\over 2}(1+z){d\over dz}\left({r^2\over (dr/dz)^2}\right)\right]
H_0^2\Omega_{k0}-{3\over 2}H_0^2\Omega_{M0}.
\eeqa
Once $\Omega_{M0}$ is measured by other observations, we can determine 
unambiguously the equation of state or effective potential of a scalar 
field from the luminosity distance $d_L(z)=(1+z)r(z)$.

\section{summary}

We have explored the possibility of determining the equation of state 
(or the metric) directly from the observational data. 
We have shown that by combining the observation of the luminosity
distance and the observation of the time variation of the cosmic
redshift, 
the  equation of state of the Universe (or the metric) 
might be directly determined unambiguously in future without any theoretical
assumption on the matter content and the geometry of the Universe. 
Since there exists no convincing candidate for the equation of state
of the x-component or the effective potential, it may be more promising to 
determine it directly from observational data.

\section*{Acknowledgments}

We would like to thank Sean Hayward for a careful reading of the
manuscript. 
This work was supported in part by a Grant-in-Aid of the Ministry of
Education, Culture, and Sports No.09640351 and 09NP0801.
One of the authors (TC) is supported by a JSPS Research Fellowship for 
Young Scientists.

\bsp

\label{lastpage}

\end{document}